\documentclass[11pt]{article}

\usepackage[utf8]{inputenc}
\usepackage[T1]{fontenc}
\usepackage{amsmath,amssymb}
\usepackage[round]{natbib}
\usepackage{hyperref}
\usepackage{orcidlink}
\usepackage[margin=1in]{geometry}
\usepackage{xcolor}
\usepackage{fancyvrb}
\usepackage{underscore}

\hypersetup{colorlinks=true, linkcolor=blue!50!black, citecolor=blue!50!black, urlcolor=blue!50!black}

\newcommand{\R}{\mathbb{R}}
\newcommand{\E}{\mathbb{E}}

\newcommand{\pkg}[1]{\textbf{#1}}
\newcommand{\proglang}[1]{\textsf{#1}}
\newcommand{\code}[1]{\texttt{#1}}

\DefineVerbatimEnvironment{CodeInput}{Verbatim}{fontsize=\small, xleftmargin=1.5em, frame=single, framesep=3pt}
\DefineVerbatimEnvironment{CodeOutput}{Verbatim}{fontsize=\small, xleftmargin=1.5em, frame=single, framesep=3pt}

\title{\pkg{corrscore}: Matrix-Aware Proper Scoring Rules and Significance
  Testing for Correlation and Covariance Forecasts in \proglang{Python}}
\author{Vinh Nguyen~\orcidlink{0009-0002-8227-5560} \\ \small Independent Researcher}
\date{\today}

\begin{document}
\maketitle

\begin{abstract}
Forecasting a correlation or covariance matrix is common in risk management and portfolio
construction, but evaluating such a forecast correctly is not routine: naive matrix-comparison
metrics are not proper scoring rules, walk-forward evaluation windows are easy to overlap with
the estimation window in ways that silently leak information, and significance testing on
serially dependent forecast-error sequences needs machinery few analysts implement from
scratch. \pkg{corrscore} is a \proglang{Python} package that provides matrix-aware
implementations of two established proper scoring rules for this setting -- the energy score
\citep{gneiting2007} and the variogram score \citep{scheuererhamill2015} -- dispatched across a
closed-form tractability spectrum (point, discrete-mixture, and isotropic-Gaussian-mixture
forecasts are scored exactly; a general Monte Carlo ensemble falls back to sampling), a
geometry-aware variant of the variogram score built from the affine-invariant distance on the
correlation manifold, a zero-overlap-by-construction walk-forward backtest harness, and a
bundled significance-testing suite (circular block bootstrap, the Diebold--Mariano test, and
the Model Confidence Set). We describe the package's design, its point of departure from the
existing \pkg{scoringRules} \citep{scoringrules2019} and \pkg{properscoring} packages, and
walk through a complete worked example.
\end{abstract}


\section{Introduction}
\label{sec:intro}

A correlation or covariance matrix forecast is the object at the center of most portfolio
construction and risk management workflows: it feeds mean-variance optimization, value-at-risk
estimation, and stress testing alike. Yet, compared to forecasting a single number, evaluating
whether one such forecast is genuinely better than another is not something the applied
literature has settled tooling for. Three specific difficulties recur in practice, each with a
known statistical fix that is nonetheless rarely implemented correctly from scratch:

\begin{enumerate}
  \item \textbf{Naive matrix-comparison metrics are usually not proper scoring rules.} A proper
    scoring rule is one under which a forecaster minimizes their own expected score precisely by
    reporting their true belief \citep{gneiting2007}; an analyst who instead compares forecasts
    by, say, mean absolute entrywise error against a single point estimate has no such guarantee,
    and worse, has no principled way to score a genuinely probabilistic (ensemble or mixture)
    forecast at all. The energy score and the variogram score are the two established proper
    scoring rules used for this purpose in the closely related field of ensemble weather
    forecasting \citep{gneiting2007,scheuererhamill2015}, but neither was published with a
    matrix-valued object, or a correlation matrix's specific structure (unit diagonal, boundary
    at $\pm 1$), in mind.
  \item \textbf{Walk-forward evaluation windows are easy to overlap by construction.} A backtest
    that estimates a forecast from data up to an origin $t$, then checks it against a
    ``realized'' correlation matrix computed from a window that reaches back before or across
    $t$, leaks information from the estimation window into the evaluation window. This is not a
    hypothetical failure mode: a window-overlap bug of exactly this kind, initially producing the
    opposite of a companion study's eventual, correctly-measured finding, motivated this
    package's backtest harness directly (Section~\ref{sec:backtest}).
  \item \textbf{Per-origin forecast scores are serially dependent, and testing them like i.i.d.\
    data overstates significance.} A walk-forward evaluation over overlapping or
    adjacent-in-time horizons produces a sequence of score differences with genuine
    autocorrelation; a plain paired $t$-test on that sequence understates its true variance.
    The Diebold--Mariano test \citep{dieboldmariano1995} and a block bootstrap
    \citep{politisromano1992} are the standard fixes, and comparing more than two candidate
    models simultaneously needs a further correction for the resulting multiple-comparison
    problem, which the Model Confidence Set procedure \citep{hansenlundenason2011} supplies.
\end{enumerate}

\pkg{corrscore} packages matrix-aware implementations of both established scoring rules, a new
geometry-aware variant of the variogram score, a backtest harness that makes the overlap bug in
item 2 structurally impossible to reproduce, and the three significance-testing tools in item 3,
so that none of this needs reimplementing per project. The package is deliberately narrow: it
scores a forecast against a realized outcome and tests whether the difference between two or
more forecasts' scores is real; it does not fit a forecasting model, and does not fetch or clean
market data.

The remainder of this paper is organized as follows. Section~\ref{sec:background} gives the
minimal proper-scoring-rule background needed to read the rest of the paper.
Section~\ref{sec:scores} describes the score implementations, including the closed-form
tractability spectrum and the geometric variogram score. Section~\ref{sec:backtest} describes
the backtest harness and the significance-testing suite. Section~\ref{sec:example} works
through a complete example. Section~\ref{sec:related} compares \pkg{corrscore} to existing
software. Section~\ref{sec:summary} concludes.

\section{Background: Proper Scoring Rules for Matrix-Valued Forecasts}
\label{sec:background}

Let $F$ be a predictive distribution over $K \times K$ correlation (or covariance) matrices, and
let $y$ be the realized outcome. A scoring rule $S(F, y)$ is \emph{proper} relative to a class
$\mathcal{P}$ of forecast distributions if, for $Y \sim G$,
$\E_G[S(G, Y)] \le \E_G[S(F, Y)]$ for all $F, G \in \mathcal{P}$ \citep{gneiting2007}: the
forecaster who reports their own true belief $G$ never does worse, in expectation, than one who
reports anything else. Both scores implemented here are instances of the general
\emph{energy-distance} recipe,
\begin{equation}
  S_\psi(F, y) \;=\; \E_F[\psi(X, y)] \;-\; \tfrac{1}{2}\,\E_F[\psi(X, X')], \qquad
  X, X' \overset{\text{iid}}{\sim} F, \label{eq:energydistance}
\end{equation}
which is proper whenever the kernel $\psi$ is \emph{conditionally negative definite}
\citep{gneiting2007,szekelyrizzo2013}. Setting $\psi$ to the Frobenius distance between two
matrices gives the \emph{energy score}; the \emph{variogram score} instead scores a vector of
pairwise absolute differences directly (Section~\ref{sec:vs}) and is proper by a separate,
simpler argument that does not require the conditional-negative-definiteness condition at all.
A companion paper \citep{nguyen2026geodesic} gives the complete propriety proofs, including for
the geometric variant of Section~\ref{sec:geovs}; this paper describes the software, not the
underlying theory, and cites that proof rather than repeating it.

\section{Score Definitions and the Closed-Form Tractability Spectrum}
\label{sec:scores}

\subsection{Forecast representation}
\label{sec:forecastrep}

Every score in \pkg{corrscore} takes a forecast as a plain \code{dict} tagged by a
\code{"kind"} key, rather than a class hierarchy, so that a caller's own forecasting code needs
no dependency on the package beyond producing this shape:

\begin{CodeInput}
{"kind": "point", "Q": Q}
{"kind": "mixture", "components": [(p_1, Q_1), ..., (p_K, Q_K)]}
{"kind": "isotropic_gaussian_mixture", "components": [(p_1, Q_1, sigma_1), ...]}
{"kind": "ensemble", "draws": [Q_1, ..., Q_M]}
\end{CodeInput}

These four kinds span a genuine closed-form tractability spectrum, not an arbitrary API choice.
A \code{"point"} forecast is a single deterministic matrix ($M = 1$). A \code{"mixture"} is any
number of discrete atoms with associated probabilities -- exactly determined, so every
expectation in Equation~\ref{eq:energydistance} reduces to a finite weighted sum, at
$O(K_{\text{atoms}}^2)$ cost, with no simulation. An \code{"isotropic_gaussian_mixture"} adds an
isotropic Gaussian scatter around each atom in the $K(K-1)/2$-dimensional space of free
(upper-triangle) matrix entries; for the energy score this remains exactly computable, via the
mean-norm formula for a noncentral chi distribution \citep{johnsonkotzbalakrishnan1994},
evaluated through the confluent hypergeometric function ${}_1F_1$ (\code{scipy.special.hyp1f1})
rather than simulated. An \code{"ensemble"} is a general Monte Carlo draw set with no assumed
structure -- the only kind with no closed form, falling back to the direct $O(M^2)$ pairwise
sum in Equation~\ref{eq:energydistance}.

\subsection{The energy score}
\label{sec:es}

\code{matrix\_energy\_score(forecast, y)} implements Equation~\ref{eq:energydistance} with
$\psi$ the Frobenius distance between two $K \times K$ matrices, dispatched across all four
forecast kinds of Section~\ref{sec:forecastrep}. For a \code{"point"} forecast this reduces to
plain Frobenius distance to the realized outcome $y$; for a \code{"mixture"} forecast with
components $(p_k, Q_k)$,
\begin{equation}
  S_{\text{ES}}(F, y) \;=\; \sum_k p_k \lVert Q_k - y \rVert_F \;-\; \tfrac{1}{2} \sum_{k,l} p_k
  p_l \lVert Q_k - Q_l \rVert_F. \label{eq:esmixture}
\end{equation}

\subsection{The variogram score}
\label{sec:vs}

\begin{CodeInput}
matrix_variogram_score(forecast, y, p=0.5, weights=None,
                        n_samples=500, random_state=None)
\end{CodeInput}
implements the variogram score
\citep{scheuererhamill2015},
\begin{equation}
  S_{\text{VS}}(F, y) \;=\; \sum_{i < j} w_{ij} \Big( |y_i - y_j|^p - \E_F|X_i - X_j|^p
  \Big)^2, \label{eq:vs}
\end{equation}
adapted to index $i, j$ over the $K(K-1)/2$ free upper-triangle entries of the correlation
matrix, rather than over the original $K$-vector the published formula was defined for: the
diagonal (always exactly $1$ for a correlation matrix) and the mirrored lower triangle both
carry no information and are excluded. \code{p} defaults to $0.5$, matching
\citet{scheuererhamill2015}'s own recommended default; \code{weights} defaults to uniform. The
score is exact (no sampling) for \code{"point"} and \code{"mixture"} forecasts, since both are
fully determined by a finite set of deterministic atoms. For \code{"ensemble"} and
\code{"isotropic\_gaussian\_mixture"} forecasts, $\E_F|X_i - X_j|^p$ has no known closed form
for general $p$ and is instead estimated by Monte Carlo with \code{n\_samples} draws (default
$500$); cost and memory scale as $O(n_{\text{samples}} \times K^2(K-1)^2/4)$, which the default
keeps modest through $K = 16$.

\subsection{A geometric variogram score}
\label{sec:geovs}

A correlation matrix is constrained to a curved region (the \emph{elliptope}), not an ordinary
Euclidean cube, and that constraint becomes increasingly binding as any entry approaches
$\pm 1$. \citet{pinsontastu2013} show the energy score is close to blind to this kind of
boundary-proximate danger for Gaussian forecasts; \citet{nguyen2026geodesic} shows the flat
variogram score of Section~\ref{sec:vs} inherits a related blind spot -- its concave power
transform saturates once two entries are already far apart, which is exactly what happens near
the boundary -- and derives a fix: \code{matrix\_geodesic\_variogram\_score} runs the identical
machinery of Equation~\ref{eq:vs}, but with every free entry $\rho$ first passed through
\begin{equation}
  \phi(\rho) \;=\; \operatorname{sign}(\rho) \sqrt{\tfrac{1}{2} \Big[ \log(1+|\rho|)^2 +
  \log(1-|\rho|)^2 \Big]}, \label{eq:phi}
\end{equation}
the signed affine-invariant (Fisher--Rao) geodesic distance from independence, treating each
entry as its own isolated $2 \times 2$ correlation matrix. $\phi$ is smooth, odd, and strictly
increasing on $(-1, 1) \to \R$, and stretches the scale specifically near $\pm 1$ where the
untransformed score has already saturated -- a close cousin of the century-old Fisher
$z$-transform $\operatorname{arctanh}(\rho)$, diverging slightly faster as $\rho \to \pm 1$.
Because Equation~\ref{eq:vs}'s propriety argument holds for \emph{any} fixed measurable
per-entry transform, not only ones with a metric interpretation \citep{nguyen2026geodesic},
\code{matrix\_geodesic\_variogram\_score} needs no separate propriety proof: it is
\code{matrix\_variogram\_score} itself, unmodified, called on $\phi$-transformed entries, and
shares that function's exact/Monte-Carlo split by forecast kind and its \code{p}/\code{weights}/
\code{n\_samples} semantics. \citet{nguyen2026geodesic} validates this construction on real,
walk-forward regime-switching correlation forecasts: the geometric variant's discrimination
advantage over the flat variogram score is real but conditional, concentrated in market episodes
where the realized correlation genuinely approaches the boundary rather than present uniformly
across an unconditional average -- we report this precisely rather than claim an unconditional
improvement, and refer the theoretically-inclined reader to that paper for the full propriety
proof, the open questions it leaves (a companion geometric energy score does not, to date,
demonstrate a comparable real-data advantage), and the complete empirical picture.

\section{Backtesting and Significance Testing}
\label{sec:backtest}

\subsection{The zero-overlap backtest harness}
\label{sec:zerooverlap}

\code{backtest\_zero\_overlap} runs a walk-forward evaluation across a sequence of forecast
origins and scores one or more named forecasting methods against a shared, model-independent
ground truth:

\begin{CodeInput}
backtest_zero_overlap(forecast_fns, ground_truth_fn, origins, horizon,
                       purge_gap=0, score_fn=matrix_energy_score,
                       severity_fn=None)
\end{CodeInput}

For each \code{origin} in \code{origins}, the harness computes ground truth from a window
starting strictly after the origin:
\begin{CodeInput}
y = ground_truth_fn(origin + 1 + purge_gap,
                     origin + 1 + purge_gap + horizon)
\end{CodeInput}
then scores every \code{forecast\_fns[name](origin)} against that same \code{y}. The specific
design choice worth calling out is what the harness
\emph{cannot} and does not attempt to police: how much history a caller's own \code{forecast\_fn}
consults internally is opaque to it (a full-history discounted filter and a short trailing
window are both permitted), exactly the same responsibility boundary
\pkg{scikit-learn}'s \code{TimeSeriesSplit} leaves to its own caller. What the harness
\emph{does} enforce, unconditionally and by construction, is that \code{ground\_truth\_fn} is
only ever called with a start point strictly after the origin. This specific guarantee exists
because of a real bug: an earlier walk-forward evaluation (in the companion regime-switching
correlation study this package grew out of) computed ``realized'' ground truth from a window
\emph{ending} near the forecast origin rather than one \emph{starting} strictly after it,
silently leaking information back into the evaluation and making a naive persistence forecast
look decisively better than a genuine forecasting model, the opposite of that study's eventual,
correctly-measured finding. \code{backtest\_zero\_overlap}'s API makes that specific mistake
structurally impossible to reproduce: the caller never controls, and never sees, the start point
passed to \code{ground\_truth\_fn}.

\subsection{Significance testing}
\label{sec:sigtest}

Three complementary significance tests operate on the per-origin score arrays a
\code{BacktestResult} returns.

\begin{CodeInput}
circular_block_bootstrap(scores_a, scores_b, block_lengths,
                          n_boot=2000, seed=None)
\end{CodeInput}
runs a percentile circular block bootstrap \citep{politisromano1992} on the paired differential
$d_i = a_i - b_i$, where $a_i, b_i$ are the $i$th entries of \code{scores\_a}, \code{scores\_b}, via \pkg{arch}'s
\code{CircularBlockBootstrap} \citep{arch2015}, swept across a list of candidate block lengths
as a sensitivity check rather than relying on one automatically ``optimal'' length; the most
conservative (largest $p$-value) entry is the one worth reporting as the headline result.

\begin{CodeInput}
diebold_mariano(loss_a, loss_b, h=1, varestimator="acf")
\end{CodeInput}
implements the
Diebold--Mariano test \citep{dieboldmariano1995} on the same paired differential, including the
Harvey--Leybourne--Newbold small-sample correction \citep{harveyleybournenewbold1997} and a
Student-$t$ (not asymptotic normal) reference distribution with $n-1$ degrees of freedom,
matching \proglang{R}'s \code{forecast::dm.test} \citep{hyndman2008forecast} formula for
formula rather than following the original 1995 paper's asymptotic-normal reference directly.
This vendored implementation is deliberately cross-checked against \code{forecast::dm.test}'s
own output on fixed synthetic data as a development-time oracle: doing so caught a genuine
normalization bug during development, where the package's long-run-variance estimator initially
normalized each sample autocovariance by $n - \text{lag}$ (the textbook ``unbiased'' convention)
rather than by the full sample size $n$ at every lag, the convention \proglang{R}'s own
\code{acf()} function -- and hence \code{forecast::dm.test} -- actually uses. The two
conventions coincide at lag $0$, so this discrepancy was invisible at horizon $h=1$ and only
surfaced once the test suite exercised $h > 1$ against the oracle.

\begin{CodeInput}
model_confidence_set(scores, alpha=0.10, block_len=5,
                      n_boot=1000, seed=None)
\end{CodeInput}
implements the Model Confidence Set \citep{hansenlundenason2011} via its range-statistic
elimination algorithm: repeatedly test whether the current candidate set of models is
statistically distinguishable from its own best member, and if so, drop the single
worst-performing model and repeat, until the surviving set cannot be rejected at level
\code{alpha}. The null distribution and each round's studentizing standard errors both come
from the same joint circular block bootstrap of the full loss matrix, reusing the identical
\pkg{arch} machinery Section~\ref{sec:sigtest}'s bootstrap function depends on. This is the
package's own reasonable operationalization of Hansen et al.'s procedure, not a line-by-line
port of any specific existing implementation's internal choices -- \proglang{R}'s \pkg{MCS}
package's automatic block-length selection, for instance, is not reproduced, and
\code{block\_len} is instead a required, caller-chosen parameter, swept manually if robustness
to it matters. The test suite accordingly cross-checks this implementation's \emph{verdict}
(which models survive) against \proglang{R}'s \code{MCS::MCSprocedure} on fixed synthetic data
where the correct verdict is unambiguous by construction, rather than requiring byte-exact
statistic or $p$-value agreement.

\section{Illustrative Example}
\label{sec:example}

The following example simulates a $K = 4$ asset panel whose true pairwise correlation drifts
slowly between roughly $0.15$ and $0.50$ over 900 trading days, and compares two deliberately
simple forecasting rules: \code{persistence}, which forecasts tomorrow's correlation matrix as
the sample correlation of the trailing 60-day window, and \code{shrinkage}, which additionally
shrinks that estimate 30\% of the way toward the identity matrix. Because the underlying process
has real, substantial, slowly-varying correlation, shrinking toward independence is the wrong
thing to do here by construction -- the example is deliberately built so that one forecast is
genuinely worse, to show what a decisive verdict from the full toolchain looks like end to end.

\begin{CodeInput}
import numpy as np
from corrscore import (backtest_zero_overlap, circular_block_bootstrap,
                        diebold_mariano, matrix_geodesic_variogram_score,
                        matrix_variogram_score, model_confidence_set)

rng = np.random.default_rng(0)
K, n_days = 4, 900
true_rho = 0.15 + 0.35 * (0.5 + 0.5 * np.sin(np.arange(n_days) / 60.0))
returns = np.empty((n_days, K))
for t in range(n_days):
    corr = np.full((K, K), true_rho[t]); np.fill_diagonal(corr, 1.0)
    returns[t] = rng.multivariate_normal(np.zeros(K), corr)

def sample_corr(start, end):
    return np.corrcoef(returns[start:end], rowvar=False)

def persistence_forecast(origin, window=60):
    return {"kind": "point", "Q": sample_corr(origin - window + 1, origin + 1)}

def shrinkage_forecast(origin, window=60, alpha=0.3):
    q = sample_corr(origin - window + 1, origin + 1)
    return {"kind": "point", "Q": (1 - alpha) * q + alpha * np.eye(K)}

origins, horizon = list(range(120, n_days - 10, 10)), 10
result = backtest_zero_overlap(
    forecast_fns={"persistence": persistence_forecast, "shrinkage": shrinkage_forecast},
    ground_truth_fn=sample_corr, origins=origins, horizon=horizon,
    score_fn=matrix_variogram_score)

print("mean VS, persistence:", result.scores["persistence"].mean())
print("mean VS, shrinkage:  ", result.scores["shrinkage"].mean())

boot = circular_block_bootstrap(result.scores["shrinkage"], result.scores["persistence"],
                                 block_lengths=[5, 10], seed=1)
for block_len, r in boot.items():
    print(f"bootstrap block_len={block_len}: mean diff={r.obs:.4f}, "
          f"95% CI=({r.ci_lo:.4f}, {r.ci_hi:.4f}), p={r.p_value:.4f}")

dm = diebold_mariano(result.scores["shrinkage"], result.scores["persistence"], h=horizon)
print(f"Diebold-Mariano: stat={dm.statistic:.3f}, p={dm.p_value:.4f}")

mcs = model_confidence_set(result.scores, alpha=0.10, block_len=10, seed=2)
print("MCS survivors:", mcs.survivors, " eliminated:", mcs.eliminated)
\end{CodeInput}
\begin{CodeOutput}
mean VS, persistence: 3.0274186774073626
mean VS, shrinkage:   3.52280921068856
bootstrap block_len=5: mean diff=0.4954, 95% CI=(0.4302, 0.5600), p=0.0000
bootstrap block_len=10: mean diff=0.4954, 95% CI=(0.4267, 0.5592), p=0.0000
Diebold-Mariano: stat=13.026, p=0.0000
MCS survivors: ['persistence']  eliminated: [('shrinkage', 0.0)]
\end{CodeOutput}

All three significance tools agree: \code{shrinkage} is decisively worse, not just
numerically worse, than \code{persistence} on this panel, and the Model Confidence Set
correctly eliminates it at $\alpha = 0.10$. Passing \code{matrix\_geodesic\_variogram\_score} as
the \code{score\_fn} argument to \code{backtest\_zero\_overlap} reuses every other line
unchanged, since Section~\ref{sec:geovs}'s geometric score shares the
flat variogram score's exact call signature -- a deliberate design consequence of implementing
it as a transform composed with existing machinery rather than a parallel code path.

\section{Comparison with Existing Software}
\label{sec:related}

\pkg{properscoring} \citep{properscoring2015} is a mature \proglang{Python} package for the
continuous ranked probability score and the Brier score, both fundamentally univariate (or
applied marginally, series by series); it has no multivariate energy or variogram score, and no
correlation-matrix-specific concept at all. \pkg{scoringRules} \citep{scoringrules2019}, an
\proglang{R} package, is the closer relative: its
\code{es\_sample} and \code{vs\_sample} functions compute the energy score and variogram score
for a general $d$-dimensional vector from a Monte Carlo ensemble sample. \pkg{corrscore}
differs in four concrete ways relevant to the correlation-matrix use case specifically: (i) a
closed-form tractability spectrum across point, discrete-mixture, and isotropic-Gaussian-mixture
forecasts (Section~\ref{sec:forecastrep}), where \pkg{scoringRules} always samples; (ii) a
$K \times K$-matrix-native API with a free-upper-triangle indexing convention built in, rather
than requiring the caller to flatten a matrix into a generic vector and manage that convention
themselves; (iii) the geometric variogram score of Section~\ref{sec:geovs}, with no analog in
either package; and (iv) a bundled walk-forward backtest harness and significance-testing suite
(Section~\ref{sec:backtest}) purpose-built around this evaluation workflow, which neither
package provides -- a \pkg{scoringRules} user must still write their own backtest loop and
reach for a separate package (e.g.\ \proglang{R}'s \pkg{MCS}) for multi-model significance
testing.

\section{Summary}
\label{sec:summary}

\pkg{corrscore} packages matrix-aware proper scoring rules, a geometry-aware variant motivated
and validated in a companion paper, and a backtest-plus-significance-testing bundle purpose-built
for evaluating correlation and covariance-matrix forecasts, closing a real, specific gap left by
existing general-purpose scoring-rule software. The package is available under the MIT license
at \url{https://github.com/vinhnguyen3455/corrscore}; 62 tests, including property-based tests
via \pkg{Hypothesis} \citep{maciver2019hypothesis} and, for the vendored Diebold--Mariano and
Model Confidence Set implementations, cross-checks against \proglang{R} reference
implementations, pass on \proglang{Python} 3.10 and later.

\bibliographystyle{plainnat}

\end{document}